\documentclass[letterpaper,twocolumn,10pt]{article}
\pdfoutput=1 
\usepackage[utf8]{inputenc}
\usepackage{subcaption}
\usepackage{epsfig}
\usepackage{diagbox}
\usepackage{adjustbox}
\usepackage{verbatim}
\usepackage{algorithm}
\usepackage{textcomp}
\usepackage{amsmath,amssymb,amsfonts}
\usepackage{graphicx}
\usepackage{algorithmicx}
\usepackage{algpseudocode}
\usepackage{xurl}
\algnewcommand\algorithmicforeach{\textbf{for each}}
\algdef{S}[FOR]{ForEach}[1]{\algorithmicforeach\ #1\ \algorithmicdo}

\algnewcommand\And{\textbf{and}}
\algrenewcommand\algorithmiccomment[2][\normalsize]{{#1\hfill\(\triangleright\) #2}}

\usepackage{multirow}

\title{Code2Image: Intelligent Code Analysis by Computer Vision Techniques and Application to Vulnerability Prediction}
\author{Zeki Bilgin\\Arcelik Research, Istanbul, Turkey \\ zeki.bilgin@arcelik.com }
\date{}

\begin{document}

\maketitle
\begin{abstract}

Intelligent code analysis has received increasing attention in parallel with the remarkable advances in the field of machine learning (ML) in recent years.  A major challenge in leveraging ML for this purpose is  to represent source code in a useful form that ML algorithms can accept as input. In this study, we present a novel method to represent source code as image while preserving semantic and syntactic properties,   which paves the way for leveraging computer vision techniques to use for code analysis. Indeed the method makes it possible to directly enter 
the resulting image representation of source codes  into deep learning (DL) algorithms  as input without requiring any further data pre-processing or feature extraction step. We demonstrate feasibility and effectiveness of our method by  realizing a vulnerability prediction use case over a public dataset containing a large number of real-world source code samples with performance evaluation in comparison to the state-of-art solutions. Our implementation is publicly available\footnote{https://github.com/ArcelikHMI/Code2Image.git}.

\end{abstract}


\section{Introduction}

Some exciting applications that fall under the scope of intelligent code analysis are vulnerability prediction \cite{Bilgin20,phdthesis, suneja2020learning}, semantic-based code search \cite{chen2019capturing},  code summarization \cite{alon2018code2seq,chen2019capturing}, code captioning \cite{alon2018code2seq}, code classification and clone detection \cite{zhang2019novel}, semantic error identification, code review and completion, synthesis, repairs, documentation and more \cite{8816775,alon2018code2seq, gupta2018intelligent}. These are mostly developed based on Artificial Intellifence (AI) in general and  Machine Learning (ML) or Deep Learning (DL) in particular.   

In traditional ML-based approaches, a significant challenge to perform intelligent  code analysis  is to extract useful features from source code, which mostly requires hand-crafted feature engineering and judicious data pre-processing steps. DL, which is a speciﬁc kind of ML,  eliminates  this difficulty by enabling  to learn features from data  automatically \cite{Goodfellow-et-al-2016}. Yet there remains an issue to be overcome in order to take advantage of DL: presenting the data in a format that DL algorithm can accept \cite{zhang2019novel}.

The high-level structure of source code is naturally text-based to facilitate the code development process which is usually fulfilled by software developers following the underlying grammar and syntax rules of the programming language. Software development processes contains useful information about the resulting code, and from this perspective,  some studies \cite{5560680, 10.1145/2635868.2635880, chernis2018machine} investigate the usefulness of several software metrics such as  code churn, developer activity, character diversity, string entropy, and  function length for code analysis. Embracing the code naturalness hypothesis that claims that code is a natural product of human effort and therefore should have common statistical properties with natural human language  \cite{hindle2016naturalness}, some other studies \cite{8816775,2018Russell} leverage natural language processing (NLP) techniques for code representation and feature extraction by treating the source code as regular text. However, treating  source code as a natural language-based text may have some drawbacks in capturing comprehensive code semantics because code is more structural and logical than natural languages \cite{zhang2019novel, NIPS2019_9209}. Therefore, some works  \cite{alon2018code2seq,10.1145/3290353,Bilgin20, chen2019capturing,  suneja2020learning,zhang2019novel} seek to develop more efficient code representation methods for intelligent code analysis, among which a promising approach is to benefit from Abstract Syntax Tree (AST) representation of source code. Due to fact that AST is a useful mid-level representation of source code and holds  rich structural and semantic information about the code \cite{chen2019capturing}, there is a spreading interest in AST-based code analysis.

Nevertheless, it is still a challenge to transfer the structural and semantic information hidden in AST to the ML/DL models  with minimal loss. In this study, we present a novel AST-based code representation method to leverage DL algorithms for intelligent code analysis. The main novelty in our approach is that the code representation we generate is in the form of image  which paves the way for leveraging AI-based computer vision techniques to use for intelligent code analysis. This is a very important achievement as the DL algorithms exhibit  spectacular successes especially in the field of computer vision. Thus, the gains achieved in the computer vision field with DL algorithms can be transferred to the intelligent code analysis area. As the saying goes, a picture is worth a thousand words. 

To demonstrate the feasibility and effectiveness of  our method, we set up a use case of vulnerability prediction from source code on a public dataset including a large number of function-level real-world code samples. Detecting software vulnerabilities before they are exposed is a highly important task to protect people and companies against malicious attempts through the exploitation of vulnerabilities \cite{9343011}. It is more difficult to detect vulnerabilities in code than to detect bugs, because vulnerabilities are often not realized by users or developers during the normal operation of the system while bugs or defects are more easily and naturally noticed \cite{phdthesis}. Moreover, vulnerability prediction should not raise high false alarms given that the rate of vulnerable code samples is very small. In our vulnerability prediction 
use case, we make performance comparison between our method and state-of-art solutions on the same dataset. Notice that although we demonstrate a vulnerability prediction use case in the scope of this study, it is possible to develop many alternative promising use cases based on our code representation method.

The main contributions of the paper are summarized as follows:

\begin{itemize}
    \item A novel visual code representation method that has the potential to leverage computer vision techniques for intelligent code analysis,
    \item Enabling automatic feature extraction from source code thanks to presented image representation method, 
    \item Converting source code's AST into a useful form while preserving syntactic and semantic information, which can directly be accepted by DL models, and 
    \item Providing a use-case of vulnerability prediction on real-world code samples in comparison with state-of-art approaches.  
\end{itemize}

The rest of the paper is organized as follows: In Section \ref{sec:relatedwork}, we overview prior studies in the areas of code representation and vulnerability prediction. Then we provide some background information about AST in Section \ref{sec:background}. We present our code representation method  in Section \ref{sec:imagerepresentation}. Section \ref{sec:vulpredict} covers our use-case of vulnerability prediction from source code using our code representation method. Finally, Section \ref{sec:conc} includes our conclusion remarks.

\section{Related Work}
\label{sec:relatedwork}

ML/DL-based intelligent code analysis, in general, relies on extracting useful features from source code. Advanced DL algorithms can extract such features automatically from code provided that the code is in a form that is directly acceptable by the algorithm, which is unfortunately not easy given the current state of technology \cite{dey2019socodecnn}. Some studies \cite{5560680, 10.1145/2635868.2635880, chernis2018machine} use certain software metrics such as complexity, code churn, character diversity, string entropy, function length, and developer activity to perform code analysis. Another group of studies leverage NLP  techniques for code representation and feature extraction. For example, the authors of \cite{2018Russell} adapt a method, which was initially developed for sentence sentiment classification,  to classify ``vulnerable'' and ``non-vulnerable'' source code components by employing deep feature representation learning over a one-hot encoding  of the tokens obtained from the lexed source code.

From a different perspective, the code2vec \cite{10.1145/3290353}  uses a neural network model to represent a code snippet as a single fixed-length code vector, by decomposing the code into a collection of paths in its AST, called path contexts, and then the network learns the atomic representation of each path contexts while simultaneously learning how to aggregate a set of them. The authors demonstrate the effectiveness of the code2vec by using it to predict a method’s (i.e. function's) name from the vector representation of its body. Although it seems that the code2vec performs well for predicting what a code fragment does, it is shown in \cite{Bilgin20} that it does not perform well for the task of vulnerability prediction. Another study benefiting from AST is  \cite{zhang2019novel}, where the authors propose an AST-based Neural Network (ASTNN) for source code representation, which  
splits each large AST into a sequence of small statement
trees, and encodes the statement trees to vectors by capturing
the lexical and syntactical knowledge of statements. Based on the
sequence of statement vectors, a bidirectional RNN model is used
to leverage the naturalness of statements and finally produce the
vector representation of a code fragment. The effectiveness of the proposed code representation is evaluated on the tasks of source code classification and code clone detection. 

In a recent study \cite{Bilgin20} , the authors propose a source code representation method that is capable of characterizing source code into a proper format for further processes in ML algorithms. The presented method in \cite{Bilgin20} converts partial AST of a given source code  into a numerical array representation while preserving structural and semantic information contained in the source code. The authors of \cite{Bilgin20} validated their code representation approach with a use case of vulnerability prediction. We make performance evaluation of our method in comparison with this study on the same dataset they used.    

The authors of \cite{allamanis2017learning} employ Gated Graph Neural Networks (GGNN) on program graphs that track the dependencies of the same variables and functions to predict variable names and detect variable misuses. Another study using the GGNN is \cite{suneja2020learning}, where source code is encoded into the Code Property Graph \cite{yamaguchi2014modeling} and then vectorized to train the GGNN for vulnerability detection. The study  \cite{dey2019socodecnn} presents a method to automatically convert  program source-codes to visual images, which could be then utilized for automated classification by convolutional neural networks (CNNs). Specifically, in   \cite{dey2019socodecnn}, an intermediate representation (IR) of code is created by the LLVM compiler, and then redundant part of the IR code (such as comments or initializations) are cleaned, and finally  ASCII value of the remaining characters are treated as a pixel value in a predefined empty image canvas. This method may not capture all semantic information in the code because the resulting image relies on the ASCII value of characters in the IR of code.

To the best of our knowledge, this is the first study that proposes a method to represent AST of source code in the form of image data that can be directly accepted by DL algorithms.     

\section{Background}
\label{sec:background}
\subsection{Abstract Syntax Tree (AST)}

\begin{figure}
    \centering
    \begin{subfigure}[b]{\linewidth}
        \centering
        \includegraphics[width=0.6\textwidth]{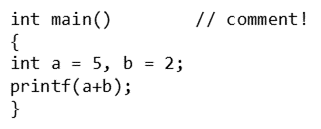}
        \caption{Source code}    
        \label{fig:source_code}
    \end{subfigure}
    \hfill
    \begin{subfigure}[b]{\linewidth}  
        \centering 
        \includegraphics[width=0.6\textwidth]{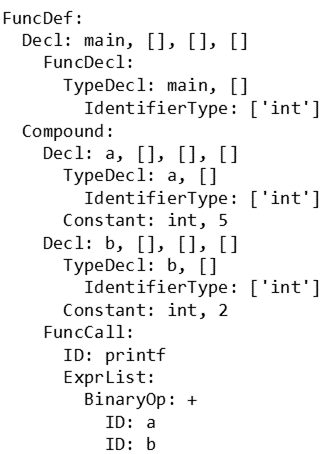}
        \caption{AST}    
        \label{fig:mainAST}
    \end{subfigure}
    \caption{A sample source code in C language  with its AST } 
    \label{fig:mainfunction}
\end{figure}

AST is a code representation in the form of a tree-type data structure. It is usually created by compilers or parsers through lexical and syntactic analysis of source code, which is a kind of tokenization and parsing process. Specifically, while generating AST representation of source code, firstly the inessential elements of code such as comments, whitespaces, tabs, and newlines are eliminated, and then the remaining part is converted into a series of tokens, where a token is a sequence of characters that can be treated as a unit in the grammar of the corresponding programming language. This can be achieved by using a lexer developed explicitly for the language of the source code.  Based on the extracted tokens, AST is generated as a tree-type graph data with hierarchical parent-child relations between the tokens. AST is an important representation of source code as it carries rich semantic and syntactic information of source code. As an example, Figure \ref{fig:mainfunction} shows a sample source code written in C language and its AST.

As seen in Figure \ref{fig:mainfunction},  while generating AST, the code is cleared of unnecessary information such as comments and tabs. Some tokens in AST shown in Figure \ref{fig:mainAST} are \textit{(FuncDef), (Decl: main, [], [], []), (IdentifierType: ['int']), (Compound), (Constant: int, 5), (FuncCall), (ID: printf), (ExprList), (BinaryOp: +), and (ID: a)}. As seen in this examples, some tokens have no parameters, while some may have one, two or three parameters. Since AST is a tree-type data, there is a parent-child relationship between tokens. For example, \textit{FuncDef} is the root, and \textit{(Compound)} is the parent of \textit{(Decl:a, [], [], []), (Decl:b, [], [], []), and (FuncCall)}. All this information (i.e. token types,  token contents, and parent-child relationships of tokens) provide rich and important information about the code.

\subsection{Code Vulnerabilities}

Vulnerability is defined as a weakness in an information system, system security procedures, internal controls, or implementation that could be exploited by a threat source \cite{Nist_800-30}, whereas a flaw or bug is a defect in a system that may (or may not) lead to a vulnerability \cite{Sate-V}. Thus, vulnerabilities are  the subclass of software bugs that can be exploited for malicious purposes \cite{dowd2006art, phdthesis}.

Detection of possible code vulnerabilities has been traditionally tried to be done by static analysis and/or   dynamic analysis. In static analysis, the code is examined for weaknesses without executing it, which does not take into account the potential impact of the executable environment such as the operating system and hardware during the analysis \cite{abraham2017review}. On the other hand, in dynamic analysis, the code is executed to check how the software will behave in a run-time environment, but this can only reason about the observed execution paths and not all possible program paths \cite{abraham2017review}. Hence, both static and dynamic code analyses have some problems on their own. Some tools used as source code security analyzer are given along with their basic capabilities by the National Institute of Standards and Technology (NIST) in the scope of Software Assurance Metrics And Tool Evaluation (SAMATE) project\footnote{\underline{https://samate.nist.gov/Main\_Page.html}}.
Given that both static and dynamic analysis may be ineffective in detecting some vulnerabilities in certain situations \cite{Sate-V, mccorkendale2014systems}, the SAMATE project presents a highly useful overview, evaluations, and test results about the effectiveness of several static code analysis tools based on a public dataset that includes real and synthetic test cases with a set of known security flaws \cite{Sate-V}.

\section{Representing Source Code as Image}
\label{sec:imagerepresentation}

We present a novel technique to generate AST-based color image representation of a given source code, which can enable to leverage deep learning-based image classification and other analysis techniques for intelligent source code analysis. The resulting AST-based image data can be readily and effectively processed, treated, and visualized as image for any intended purpose of software analysis.

\subsection{Image Representation}

As mentioned in the previous section, AST contains rich information about source code in the form of tokens and their parent-child relations. It is important to  preserve all this information in AST while generating its   image representation. To achieve this, we exploit figural,  chromatic, and spatial properties of the image as follows:

\begin{itemize}
    \item  \textbf{(figural)} We plot the tokens in a particular shape, preferred as rectangular in our implementation but it could be any other shape as well,
    \item \textbf{(chromatic)} We use colors to encode the content of AST tokens into image   such that the rectangular token shapes are filled with  specific colors predetermined depending on the type and content of the tokens, and
    \item \textbf{(spatial)} We use black lines connecting parent-child tokens such that these lines do not cross each other as in tree-based graphs.
\end{itemize}

\begin{figure}[h]
\centering
\includegraphics[width=\linewidth]{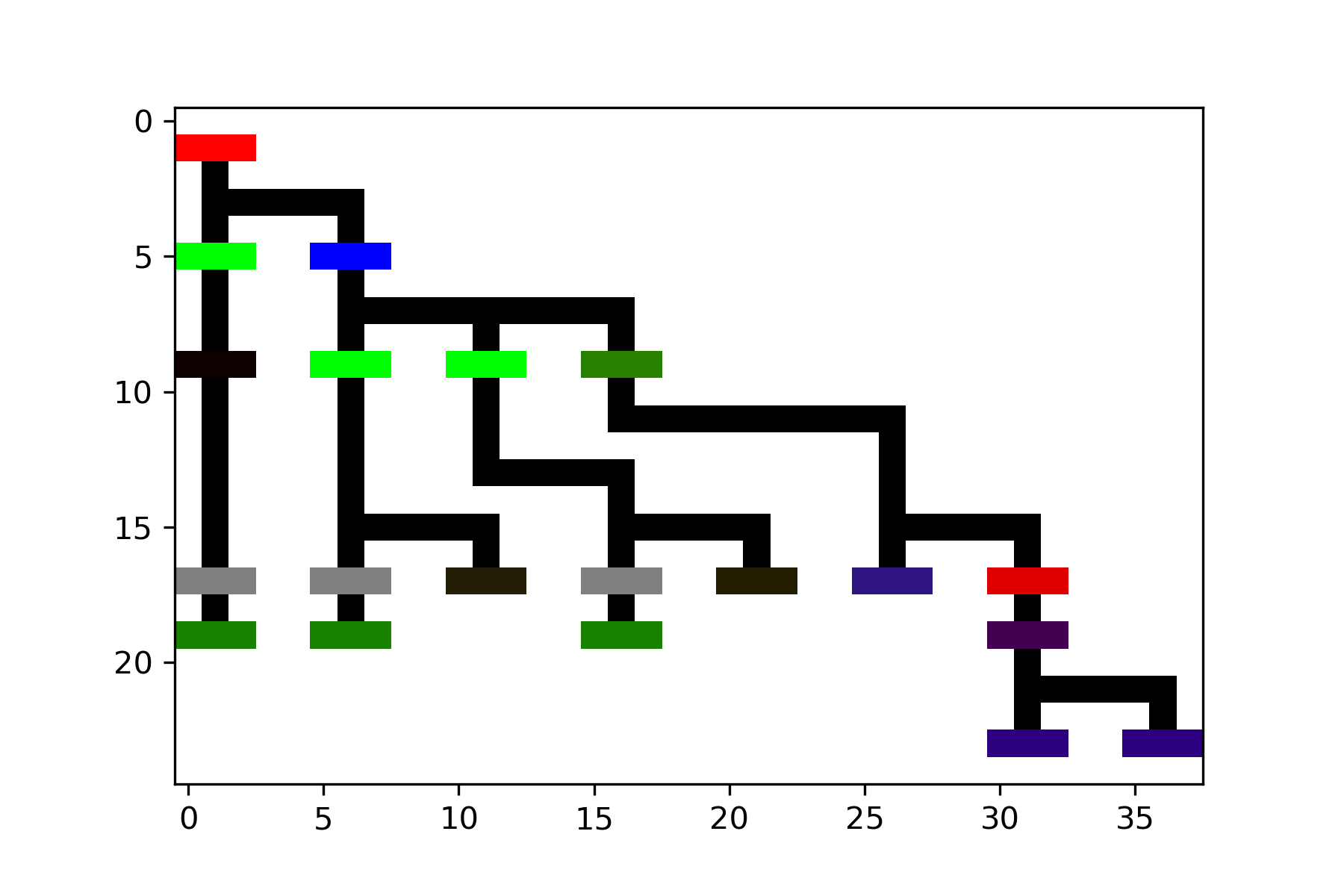}
\caption{AST-based visual image representation of the code given in Figure \ref{fig:source_code}}
\label{fig:imagerepresentation}
\end{figure}

To better explain this with a concrete example, we generate an image representation of the code given in Figure \ref{fig:source_code} by our method and visualized it in Figure \ref{fig:imagerepresentation}. 
The image in Figure \ref{fig:imagerepresentation} represents AST of the aforementioned code sample such that the rectangle boxes represent the AST tokens and are colored according to the contents of tokens.  The black lines in the image indicate the parent-child relationship between the tokens. Notice that the generated image representation  is actually in the form of RGB image data composed of 8-bit unsigned integers specifying the color of each pixel. Figure  \ref{fig:imagerepresentation} is the visualization of resulting image data, where the numbers on the x-axis and y-axis are indexes of pixels. As illustrated in this example, the original text-based AST of the code is represented in image format ready for further processing such as DL-based image classification tasks.

\begin{figure}[h]
\centering
\includegraphics[width=0.75\linewidth]{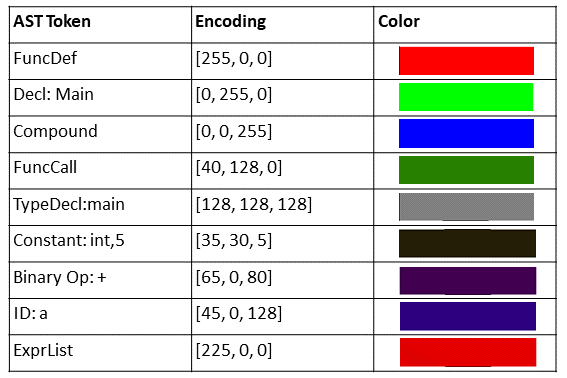}
\caption{Some examples for encoding and color mapping of AST tokens }
\label{fig:colormap}
\end{figure}

More specifically, a digital color image contains color information for each pixel. Typically, colors in an RGB image are specified by a tuple of three integer values between 0 and 255, which indicate the intensity and chroma (color) of  light. As mentioned in the previous part, an AST token may have one, two, or three content parameters. Thus, these contents can be mapped to three color values while  generating the image representation of an AST. For example, as shown in Figure \ref{fig:colormap}, we encode the token of \textit{FuncDef} as [255, 0, 0] which corresponds to a specific shade of \textit{Red} color. Similarly, other token types can be encoded to a different triplet of integers, between 0 and 255, corresponding to different colors as exemplified in Figure \ref{fig:colormap}. Notice that the colors corresponding to AST tokens can be chosen as desired provided that they are consistent across the entire dataset.    

To generate this kind of image representation accurately and automatically for any code sample, it is necessary to overcome some technical difficulties during the implementation phase as explained in the following parts.          

\subsection{Image Size}
\label{sec:imagesize}

Trees are planar graphs that can be embedded in the plane. In other words, a tree can  be drawn on the plane in such a way that its edges intersect only at their endpoints \cite{trudeau2013introduction}, which requires us to draw AST in our image representation in such a way that no edges cross each other. It is a challenge to achieve this while automatically drawing ASTs of a large number of code samples in a bounded drawing space because ASTs may have arbitrary depth length and  number of nodes. Put it another way, we draw AST by coloring associated pixels in an image canvas, and even when we set the line (i.e. edge) thickness as the minimum value of 1 pixel, the number of edges that can pass through between the nodes (i.e. rectangular color boxes in our case) without overlapping is bounded by the space between nodes. To overcome this issue,  we need to leave enough length of space between rectangular token boxes in both horizontal and vertical directions while generating the image representation of source code. But, what would be the optimal length of such a blank space? If the space becomes unnecessarily long, the resulting image sizes would be needlessly big, increasing space and processing requirements. On the other hand, if this space becomes not long enough, the black connecting lines may intersect in the resulting image representation for some source code samples.        

We develop a 2-step process to overcome this issue as follows:

\begin{enumerate}
    \item \textbf{(Drawing)} We start drawing the image from the upper left corner of the drawing canvas (i.e. 3d array initialized with the pixel value of white color)  by adjusting the space between the rectangular boxes of tokens  to the longest distance possible to avoid edge intersections, and  
    \item \textbf{(Compacting)} We make the produced image as compact as possible by removing inessential parts such as all-white columns $\&$ rows of pixels and recurring rows of pixels. 
\end{enumerate}

We explain these steps in detail in the following subsections.

\subsubsection{Drawing}
\label{subsection_drawing}

To determine the upper limit of the number of edges passing through between the token boxes in the image representation, we  performed some statistical analysis on a large number of real-world code samples. For this purpose, we benefited from the public Draper VDISC Dataset\footnote{\underline{https://osf.io/d45bw/}} \cite{2018Russell} that consists of more than 1 million function-level C and C++ code samples mined from real-world projects such as Debian Linux distribution \cite{Debian}, public Git repositories on GitHub \cite{Github}, and SATE IV Juliet Test Suite \cite{black2018juliet}. 

\begin{figure}[t]
\centering
\begin{subfigure}{1.0\linewidth}
  \centering
\includegraphics[width=\linewidth]{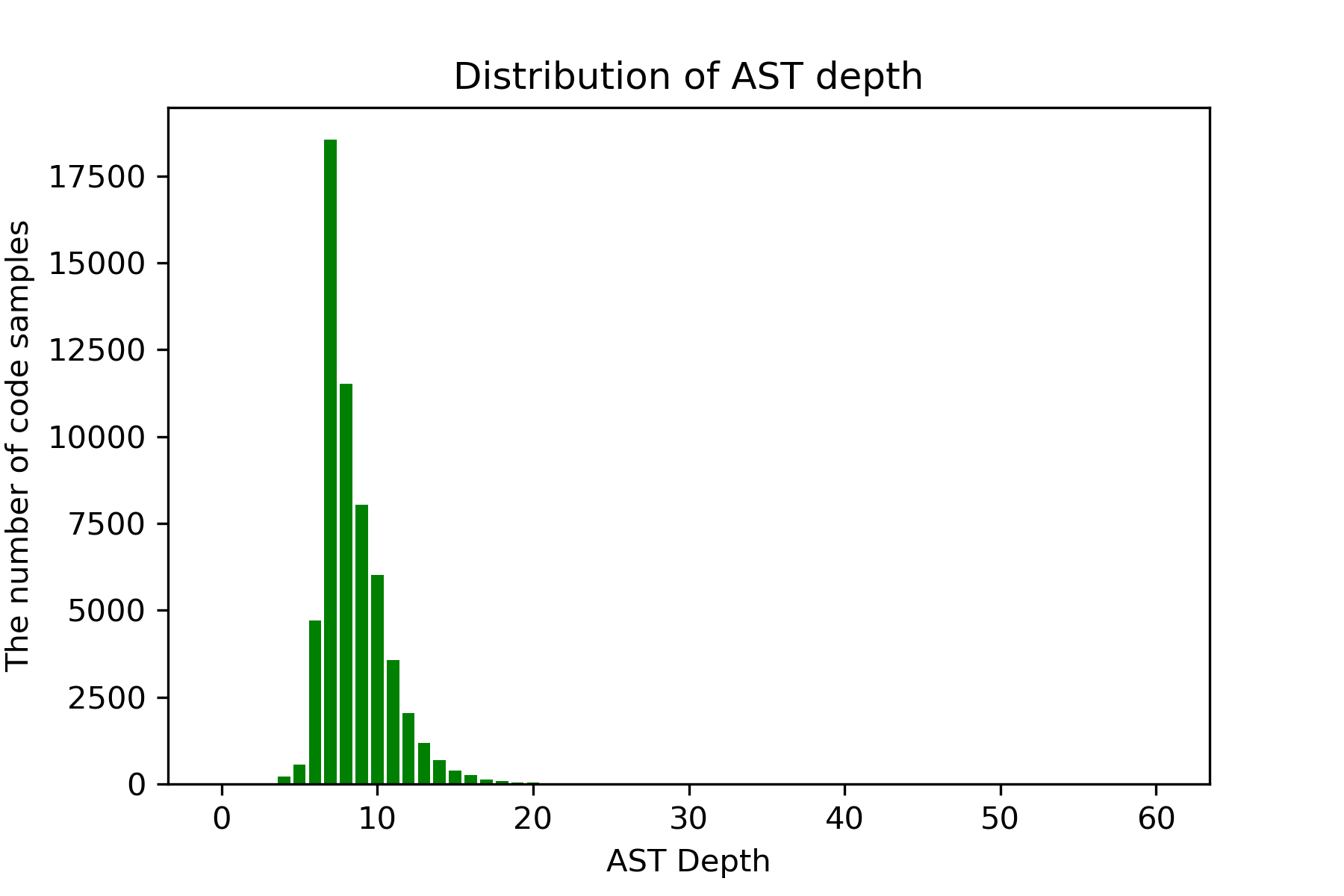}
\caption{}
\label{fig:depthdistribution}
\end{subfigure}
\begin{subfigure}{1.0\linewidth}
  \centering
\includegraphics[width=\linewidth]{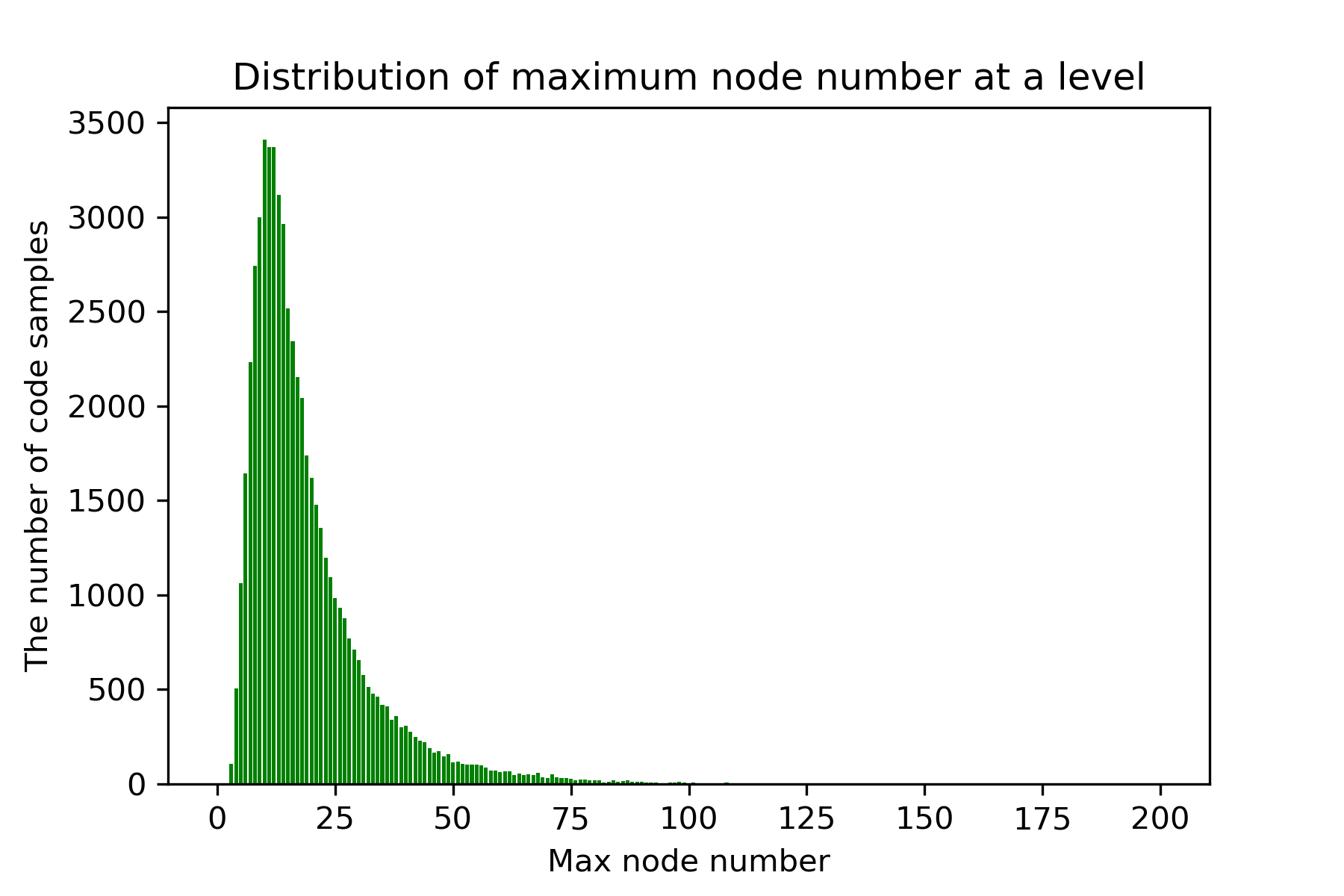}
\caption{}
\label{fig:maxnodes}
\end{subfigure}
\caption{Statistical  distributions of some AST properties}
\label{fig:ASTdistribution}
\end{figure}

In our statistical analyses, we examined approximately 50000 function-level code samples written in C language from the Draper dataset. We first extracted ASTs of those code samples by using the Pycparser\cite{pycparser}, which is a Python-based parser for the C language (C99), and explored this ASTs' structure. We noted both AST depth (i.e. level of the deepest node) and the number of nodes (i.e. tokens) at each level of ASTs to reveal the expected size of ASTs.  Figure \ref{fig:ASTdistribution} shows the statistical  distribution of these features for the code samples we investigated. According to Figure \ref{fig:depthdistribution}, the most frequent AST depths encountered are 8, 9, 10, 11, 7, 12, 13, and so on from most frequent to least. On the other hand, Figure \ref{fig:maxnodes} shows the statistical distribution of the maximum number of nodes seen at an AST level, which indicates that the highest number of nodes at AST levels is 11 for most code samples and  gradually decreases for other values. The maximum highest node number seen is 180 in just one sample. Taking into account these statistical values, we optimize our implementation by leaving enough space both horizontally between nodes at the same level and vertically between subsequent levels.

\subsubsection{Compacting}

\begin{figure}[t]
\centering
\includegraphics[width=\linewidth]{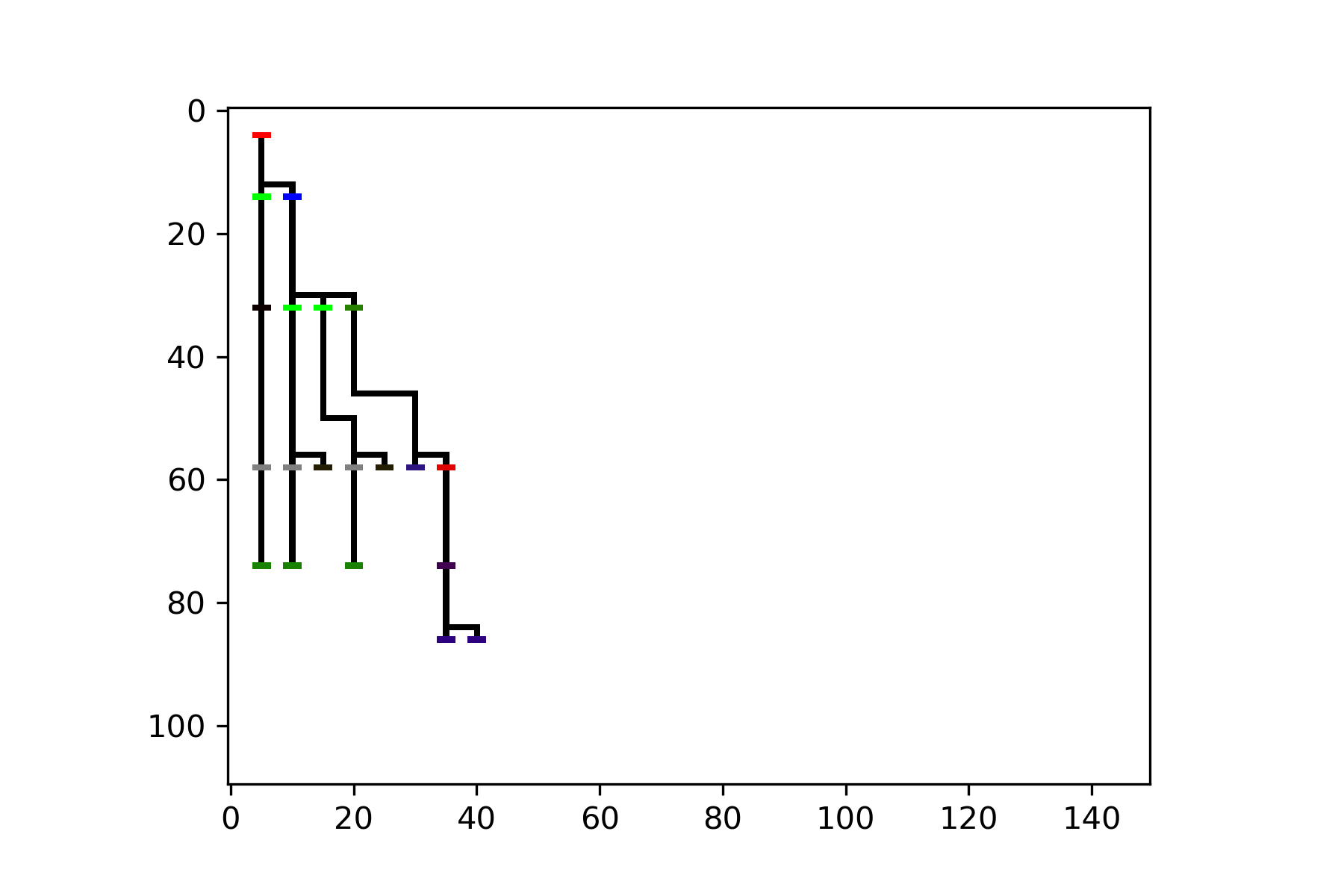}
\caption{Illustration of the impact of the compacting step. The image representation given in Figure \ref{fig:imagerepresentation} would look like this  if the compacting step was not applied.  }
\label{fig:nocompacting}
\end{figure}

When the blank space left between the rectangular token boxes is more than necessary, inessential rows and columns of pixels may occur in the resulting image representation. Specifically, there may be rows of pixels that sequentially repeats themselves, which would  unnecessarily lengthen the height of the image. Also, there may occur all-white (i.e. blank) columns and rows of pixels on the right and bottom side of the image, which would  unnecessarily increase the image size. The resulting image can be compacted by deleting this kind of inessential parts. Figure \ref{fig:nocompacting} illustrates the impact of compacting step by depicting the state of the image representation given in Figure \ref{fig:imagerepresentation} when  the compacting step is not applied. When we compare  Figure \ref{fig:imagerepresentation} and Figure \ref{fig:nocompacting}, it is seen that the compacting step brings the resulting image representation to the smallest possible size. Notice that size of the rectangular token boxes and thickness of the edges in both drawings are the same in fact but appear large in one and small in the other due to the scale difference as seen on the numbers on x-axis and y-axis.

Our implementation is summarized in the pseudocode given in Algorithm \ref{pseudocode}.

\begin{algorithm}[t]
\caption{Generating Image Representation}\label{algorithm}
\footnotesize
\label{pseudocode}
\hspace*{\algorithmicindent} \textbf{Input:} AST \hspace*{\algorithmicindent} \Comment[\footnotesize]{AST of source code } \\
\hspace*{\algorithmicindent} \textbf{Output:} Image Representation   
\begin{algorithmic}[1]
\Procedure{Encode}{node} \hspace*{\algorithmicindent} \Comment[\footnotesize]{specify colour based on token}
\State $colour \gets node$
\State \Return $colour$
\EndProcedure
\Procedure{Drawing}{root,x,y,3} \hspace*{\algorithmicindent}  \Comment[\footnotesize]{specify image dimension}
\State $image \gets  array[x][y]= [255,255,255]$ \hspace*{\algorithmicindent}  \Comment[\footnotesize]{initialize: all pixels are white}
\State $queue \gets root$  
\State $level \gets 0$
\While {$queue \neq \emptyset$}
\State $node \gets queue.pop(0)$ 
\State $level \gets node.level$
\State $index \gets placeholder$
\If {$currentlevel == level$} \hspace*{\algorithmicindent}  \Comment[\footnotesize]{draw token box}
\State $image[currentlevel][index] \gets \Call{encode}{node}$
\Else
\State $level \gets currentlevel$
\State $image[currentlevel][index] \gets \Call{encode}{node}$
\EndIf 
\If {$currentlevel \neq 0$} \hspace*{\algorithmicindent}  \Comment[\footnotesize]{draw line from child to parent}
\ForEach {$(index_i, index_j) \in line $}
\State $image[index_i][index_j] \gets 0 $   \hspace*{\algorithmicindent}  \Comment[\footnotesize]{colour black}
\EndFor
\EndIf
\State  $queue \gets nextnode$ 
\EndWhile
\EndProcedure
\Procedure{Compact}{image} 
\State $image \gets image.delete(subsequently$ $repeating$ $rows)$
\State $image \gets image.delete(allwhite$ $columns\&rows)$
\EndProcedure
\end{algorithmic}
\end{algorithm}

\subsubsection{Image Format}

The generated image representation is actually in the form of 3-dimensional array, where each dimension corresponds to 1 color channel as explained previously. This data can be converted to any intended image format such as jpeg, eps, png. Either raw form or converted form can then be directly used for further processes in machine learning applications.

\section{Vulnerability Prediction}
\label{sec:vulpredict}

We claim that our method of generating the image representation of source code enables us to leverage AI-based computer vision techniques for intelligent code analysis. To validate this, we apply our code representation method for the use case of vulnerability prediction from source code and make performance evaluation in comparison with state-of-art solutions.  

\subsection{DataSet}

The public Draper VDISC Dataset \cite{2018Russell} provides a large number of useful function-level real-world code samples that are labeled 
according to whether they contain any of the certain vulnerabilities based on the examinations conducted by several static code analyzers. The authors of \cite{2018Russell} also split the whole dataset into 3 disjoint subsets as training, validation, and test sets with the percentage of 80, 10, and 10 respectively. From this dataset, we extracted the code samples  parseable by  the Pycparser\cite{pycparser}, a Python-based parser for the C language (C99).

\begin{table*}[!tp]
	\begin{center}
	    \caption{Types and frequencies of the vulnerabilities investigated in the experimental work}
	    \resizebox{\textwidth}{!}{
		\begin{tabular}{llr} 
		\hline
			\textbf{CWE ID } & \textbf{CWE Description}  & \textbf{Frequency (\%)} \\
			\hline
			119 & Improper Restriction of Operators within the Bounds of a Memory Buffer & 21.34\\
			120/121/122 & Buffer Overflow & 40.69\\
			469 & Use of Pointer Subtraction to Determine Size & 2.57\\
			476 & NULL Pointer Dereference & 9.22\\
			20,457,805 etc & Improper Input Validation, Use of Uninitialized Variable, Buffer Access with Incorrect Length Value etc. & 26.18 \\
			\hline
		\end{tabular}}
	\label{table_CWE_description}
	\end{center}
\end{table*}

Table \ref{table_CWE_description} shows the details of the 5 vulnerability categories considered in this study. The first, third, and fourth categories contain only one type of vulnerability, which is CWE-119, CWE-469, and CWE-476 respectively. The second category includes three variants of buffer overflow, which are CWE-120 (Classic Buffer Overflow), CWE-121 (Stack-based Buffer Overflow) and CWE-122 (Heap-based Buffer Overflow). The fifth category includes a few variants of improper input validation such as CWE-20 (Improper Input Validation), CWE-457 (Use of Uninitialized Variable), and CWE-805 (Buffer Access with Incorrect Length Value). These are serious vulnerabilities that can cause irreparable damage to system, businesses, or users, and therefore it is crucial to detect them  as early as possible.

\subsection{Experimental Setup}

To realize the vulnerability prediction use case, we first generated the image representation of the code samples that we extracted from the Draper dataset. As explained earlier, we followed a two-step process for this purpose, i.e. (i) drawing and (ii) compacting. In the drawing step, to avoid the intersection of black lines connecting parent and child nodes in AST, we left long enough space horizontally and vertically between rectangular token boxes. We found the optimal length of these spaces for the whole dataset based on our statistical analysis of the ASTs extracted from a large number of real-world code samples as explained earlier in Section \ref{subsection_drawing}. In the compacting step, we removed inessential columns and rows in the resulting image, and  thus compacted them. The Appendix includes more code samples along with their AST and resulting image representations.

\subsubsection{Problem Modeling} 

We model the vulnerability prediction task as a binary image classification problem such that we train our DL-based model with the training data including code samples that are labeled either as vulnerable or non-vulnerable. The model learns  how to distinguish vulnerable and non-vulnerable code samples from training data and then we make performance evaluation on test dataset including both vulnerable and non-vulnerable code samples. We train a separate DL model for each of the 5 different vulnerability types, the details of which are given in Table  \ref{table_CWE_description}.

\subsubsection{Model Architecture}

As highlighted earlier,  DL algorithms demonstrate spectacular performance in image classification, which is mostly done by using convolutional neural networks (CNNs). We can leverage this for vulnerability prediction in source code by using  our code representation method which converts a given code fragment into image data. However, it is known that traditional CNN models accept only images of the same size due to fully connected (dense) layers, which is  a requirement that is not satisfied in our case. As discussed in Section \ref{sec:imagesize}, the ASTs of code samples collected from real-world projects have varying dimensions, and as a result of this, our method generates images in varying dimensions. 
To overcome this challenge, we initially tried to use a fully convolutional neural network (FCN) that does not contain any \textit{dense} layer, and thus accepts varying input sizes. However, we observed that the classification performance of FCN was not satisfactory. Then we modified our  convolutional network to include a dense layer with the objective of achieving better classification performance and found  another  way to solve the issue of varying input size as explained below.  

In the training process of a neural network, all samples in the training dataset are not processed at once. Instead, they are processed in batches containing a certain number of samples to reduce memory requirements.  To accept images in varying size while there is a fully connected layer in our network, we perform batch-based size equalization such that we first find the max height and width of images in a batch and pad every other image with a pixel value of 255 (i.e. white color in our case) in the right and down directions so that every image in the batch has an equal dimension. The padded white pixels in the image have no impact on code semantic because they are identical with the white background of the image. The model can 
extract features from the intended portion from the padded image. Thus we have a batch with equal image dimensions but every batch has a different shape (due to difference in max height and width of images across batches) \cite{FCN}. 

\begin{figure}[!b]
\centering 
  \includegraphics[width=\linewidth]{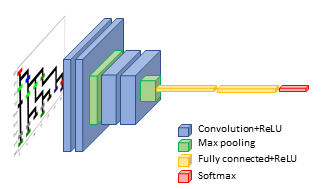}
  \caption{Architecture of our convolutional neural network}
  \label{fig:CN_architecture}
\end{figure}


\begin{table*}[hbt!]
	\caption{Distribution of positive and negative samples in the original and oversampled cases }
\label{table-imbalanced}
	\centering
	\resizebox{\textwidth}{!}{
	\begin{tabular}{l |c |c |c |c |c |c |c }
	\hline
	\hline
		\multicolumn{2}{c|}{Class} & \multicolumn{2}{c|}{Training (\# of samples:46093)} & \multicolumn{2}{c|}{Validation(\# of samples:5736)} & \multicolumn{2}{c}{Test(\# of samples:5765)}\\
		\hline
		& & Vulnerable & Non-Vulnerable & Vulnerable & Non-Vulnerable & Vulnerable & Non-Vulnerable\\
		\hline
		\multirow{2}{*}{CWE119} & original & 2684 (5.82\%) & 43409 (94.18\%) & 335 (5.84\%) & 5401 (94.16\%) & 355 (6.16\%) & 5410 (93.84\%)\\
		\cline{2-8}
		 & oversampled & 14470 (25.00\%) & 43409 (75.00\%) & - & - & - & -\\
		\hline
		\multirow{2}{*}{CWE120} & original & 5119 (11.10\%) & 40974 (88.89\%) & 641 (11.18\%) & 5095 (88.82\%) & 684 (11.86\%) & 5081 (88.14\%)\\
		\cline{2-8}
		 & oversampled & 13658 (25.00\%) & 40974 (75.00\%) & - & - & - & -\\
		\hline
		\multirow{2}{*}{CWE469} & original & 323 (0.70\%) & 45770 (99.30\%) & 36 (0.63\%) & 5700 (99.37\%) & 32 (0.56\%) & 5733 (99.44\%)\\
		\cline{2-8}
		 & oversampled & 15257 (25.00\%) & 45770 (75.00\%) &  - & - & - & -\\
		\hline
		\multirow{2}{*}{CWE476} & original & 1160 (2.52\%) & 44933 (97.48\%) & 146 (2.55\%) & 5590 (97.45\%) & 140 (2.43\%) & 5625 (97.57\%)\\
		\cline{2-8}
		 & oversampled & 14978 (25.00\%) & 44933 (75.00\%) &  - & - & - & -\\
		\hline
		\multirow{2}{*}{CWEOther} & original & 3294 (7.15\%) & 42799 (92.85\%) & 419 (7.30\%) & 5317 (92.70\%) & 399 (6.92\%) & 5366 (93.08\%)\\
		\cline{2-8}
		 & oversampled & 14266 (25.00\%) & 42799 (75.00\%) &  - & - & - & -\\
		\hline
		\hline
	\end{tabular}}
\end{table*}

In our convolutional network, we specify input shape as \textit{(None, None, 3)} since the dimensions of our input images are variable. The 3 is for the number of channels in our image which is fixed for colored images (RGB). As illustrated in Figure   \ref{fig:CN_architecture}, there are two convolution blocks in our network, each of which  consists of a 2D convolution layer \textit{(Conv2D)} with 32 filters and kernel size of 3 to extract features from input image and  an activation layer \textit{(Relu)} to  incorporate non-linearity. To avoid overfitting we also added regularization layer \textit{(Dropout and BatchNormalization)} after \textit{Conv2D}. There is a max pool layer after each convolution block, and then a full connected layer (size of 32) with an activation layer to perform the classification process. Finally, there is a softmax layer to calculate class probabilities for the corresponding input. 

\subsubsection{Dealing With Data Imbalance}

The  Draper VDISC Dataset is highly imbalanced as the number of positive (i.e. vulnerable)  samples is far less than the number of negative (non-vulnerable) samples due to fact that they are collected from real-world projects  and thus reflect the natural distribution of the targeted vulnerabilities. It is inherently challenging to deal with classification problems on highly imbalanced sets because they require to accurately detect positive test samples constituting  a very small portion of the whole test samples without raising false alarms for negative samples. When an ML/DL algorithm is trained with such an imbalanced dataset,  the impact of minority class over the trained model is damped in general due to the dominance of the majority class. As a result of this, the model can not learn enough from minority class samples, which degrades the classification performance in terms of $Precision$, $Recall$, and other metrics. As a remedy for this, we apply the oversampling technique to increase the number of vulnerable code samples only in the training dataset while not touching the test dataset. To do this, we create multiple copies of vulnerable code samples to soften the imbalance rate between classes. In this way, we adjusted the ratio of vulnerable code samples to 25$\%$ in the training dataset for all vulnerability types. Table \ref{table-imbalanced} shows the number of vulnerable and non-vulnerable code samples before and after oversampling process.  

\subsection{Experimental Results}

We trained a separate DL model for each vulnerability type (i.e. totally 5 models) on the oversampled training data to perform binary classification, and made performance evaluation on the test dataset given in  Table \ref{table-imbalanced}. We observed the performance of our model for each epoch during the training process and saved the model state whenever a lower validation loss is obtained.  We used the model state with minimum validation loss for performance evaluation on the test dataset. Also, we compared the performance of our vulnerability prediction approach with the Ast2vec \cite{Bilgin20} on the same dataset. The Ast2vec is a state-of-art code representation and vulnerability prediction  method as shortly explained in Section \ref{sec:relatedwork}.

We preferred to use F1, MMC, and AUC as performance metrics because our dataset is highly imbalanced.  Notice that our classification models are naturally probabilistic, which means they actually return probabilities indicating the likelihood of a sample belonging to each class label (i.e. vulnerable or non-vulnerable). These probabilities are converted to class labels based on decision threshold, i.e., a value above that threshold indicates ``vulnerable''; a value below indicates ``non-vulnerable''. 
A model can be operated at the optimum point (usually where the F1 score is maximum) by adjusting the decision threshold. The precision-recall curves below were obtained in this way by varying decision threshold and observing model performance in terms of precision and recall.

\begin{figure*}[htb]
    \centering 
\begin{subfigure}{0.33\textwidth}
  \includegraphics[width=\linewidth]{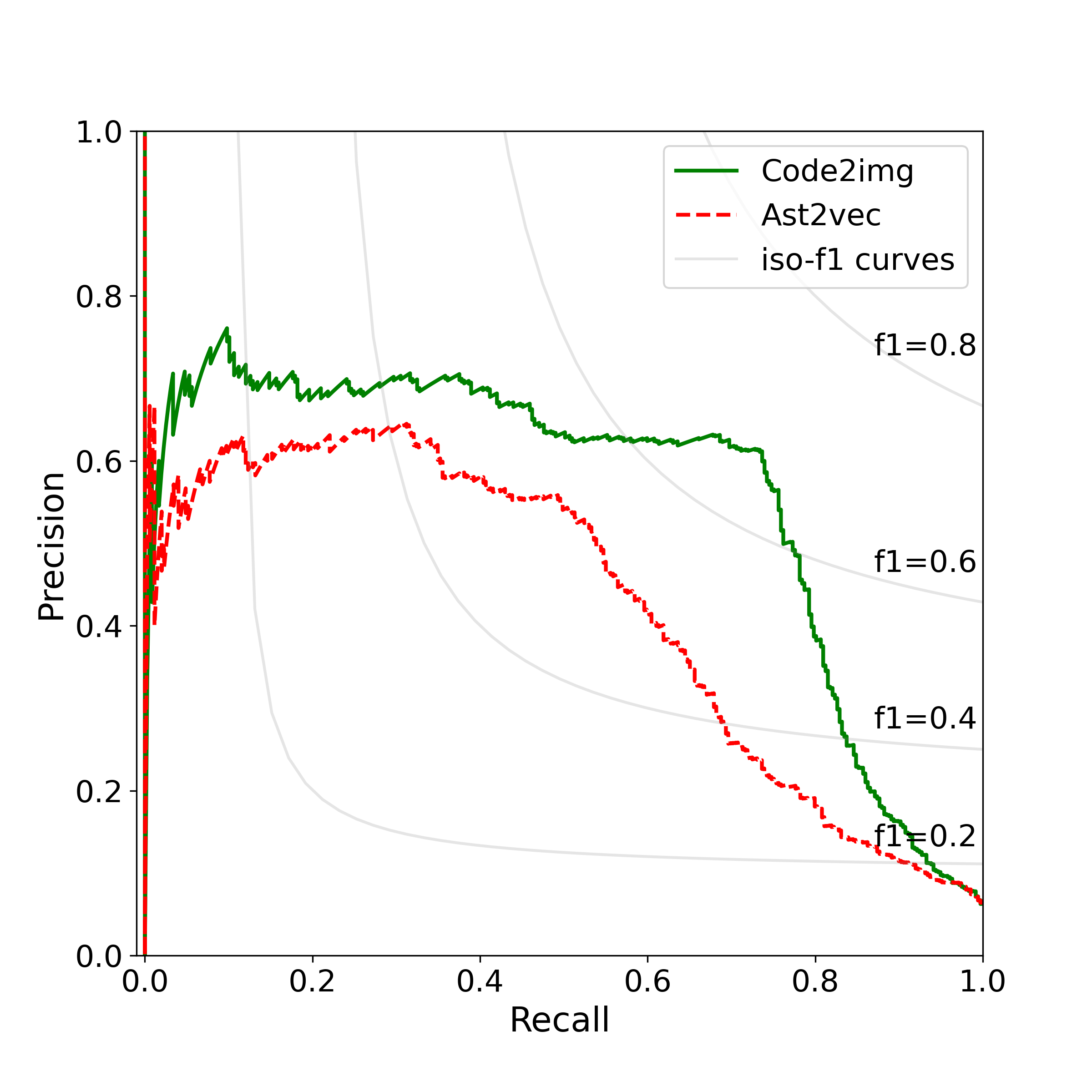}
  \caption{CWE-119}
  \label{fig:CWE119}
\end{subfigure}\hfil 
\begin{subfigure}{0.33\textwidth}
  \includegraphics[width=\linewidth]{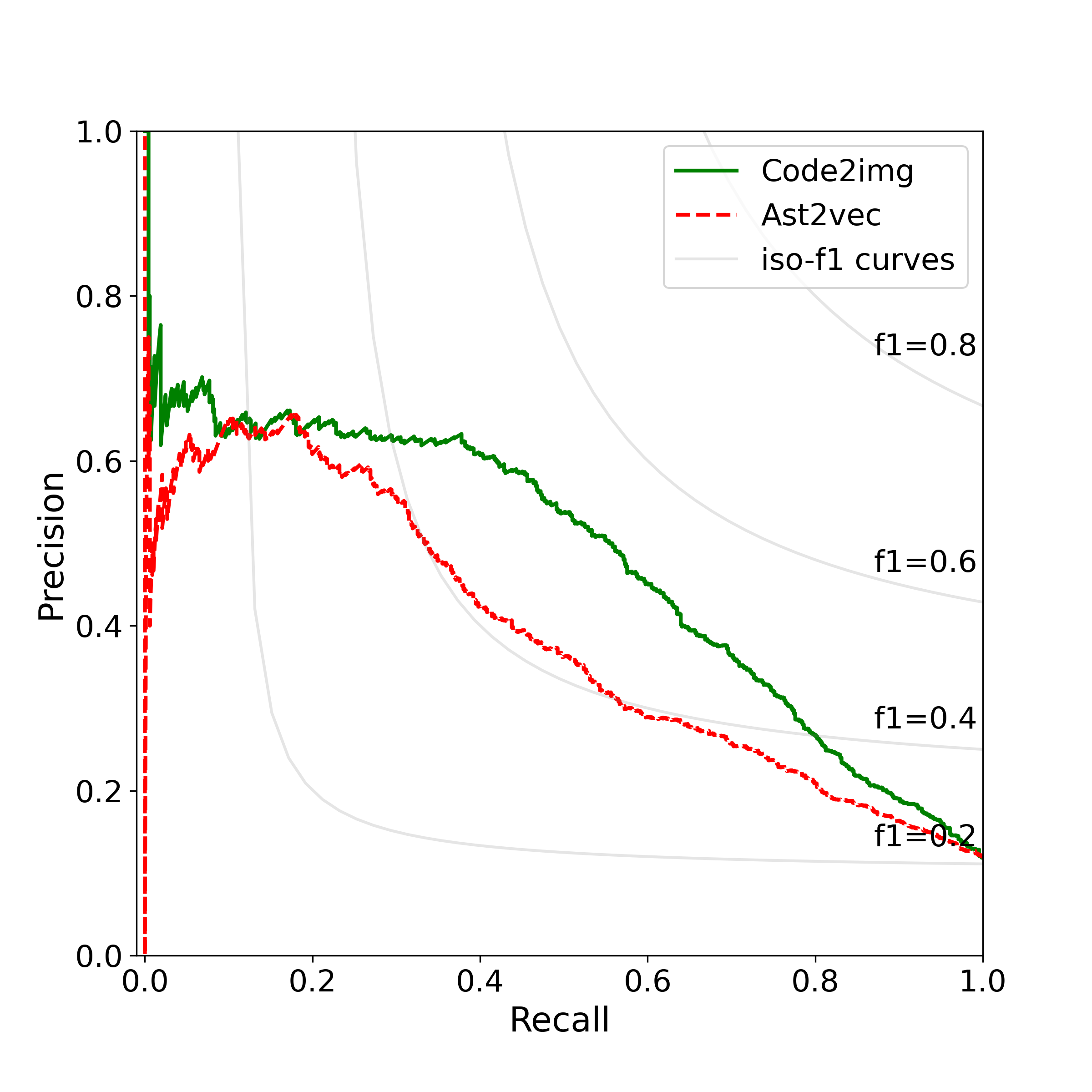}
  \caption{CWE-120}
  \label{fig:CWE120}
\end{subfigure}\hfil 
\begin{subfigure}{0.33\textwidth}
  \includegraphics[width=\linewidth]{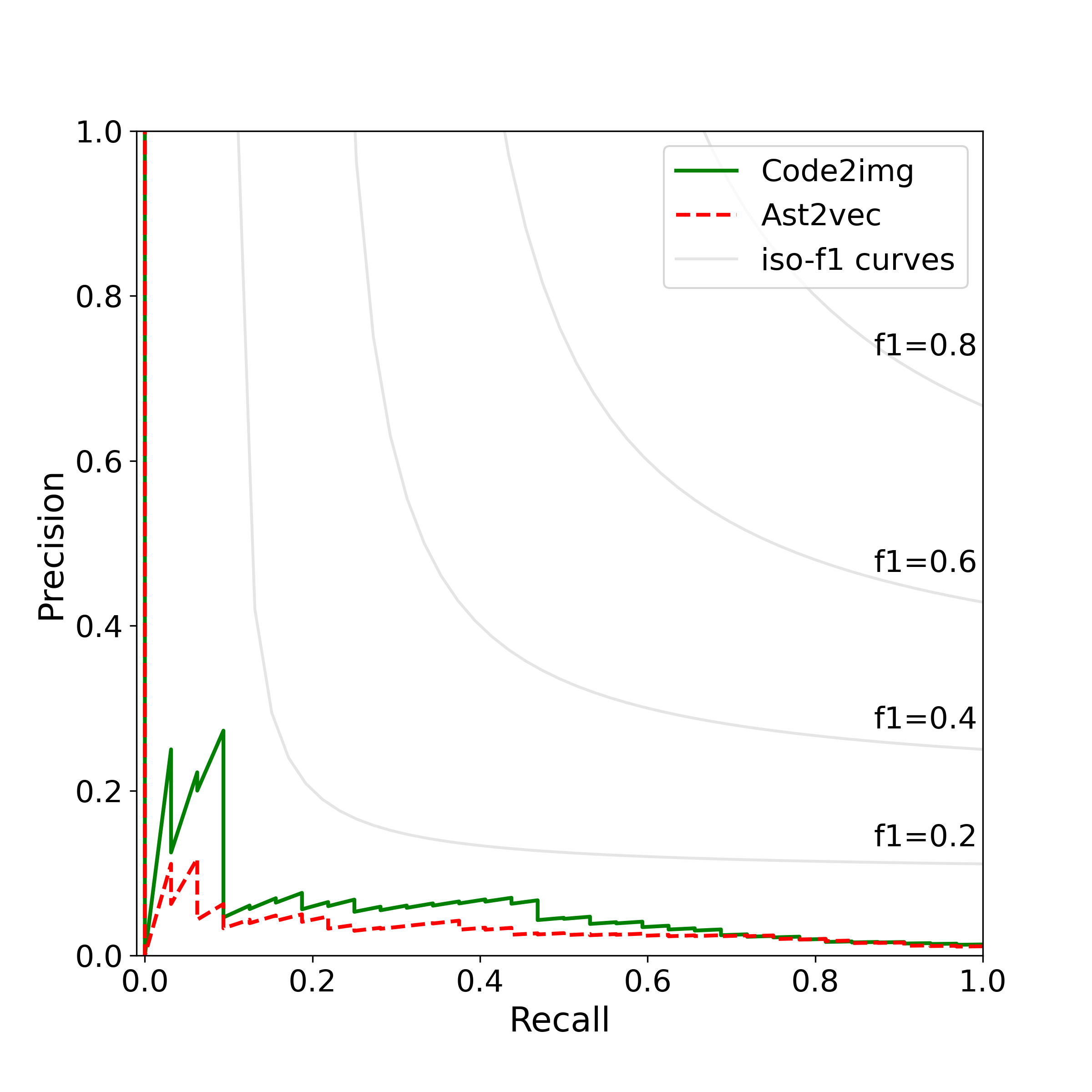}
  \caption{CWE-469}
  \label{fig:CWE469}
\end{subfigure}
\medskip
\begin{subfigure}{0.33\textwidth}
  \includegraphics[width=\linewidth]{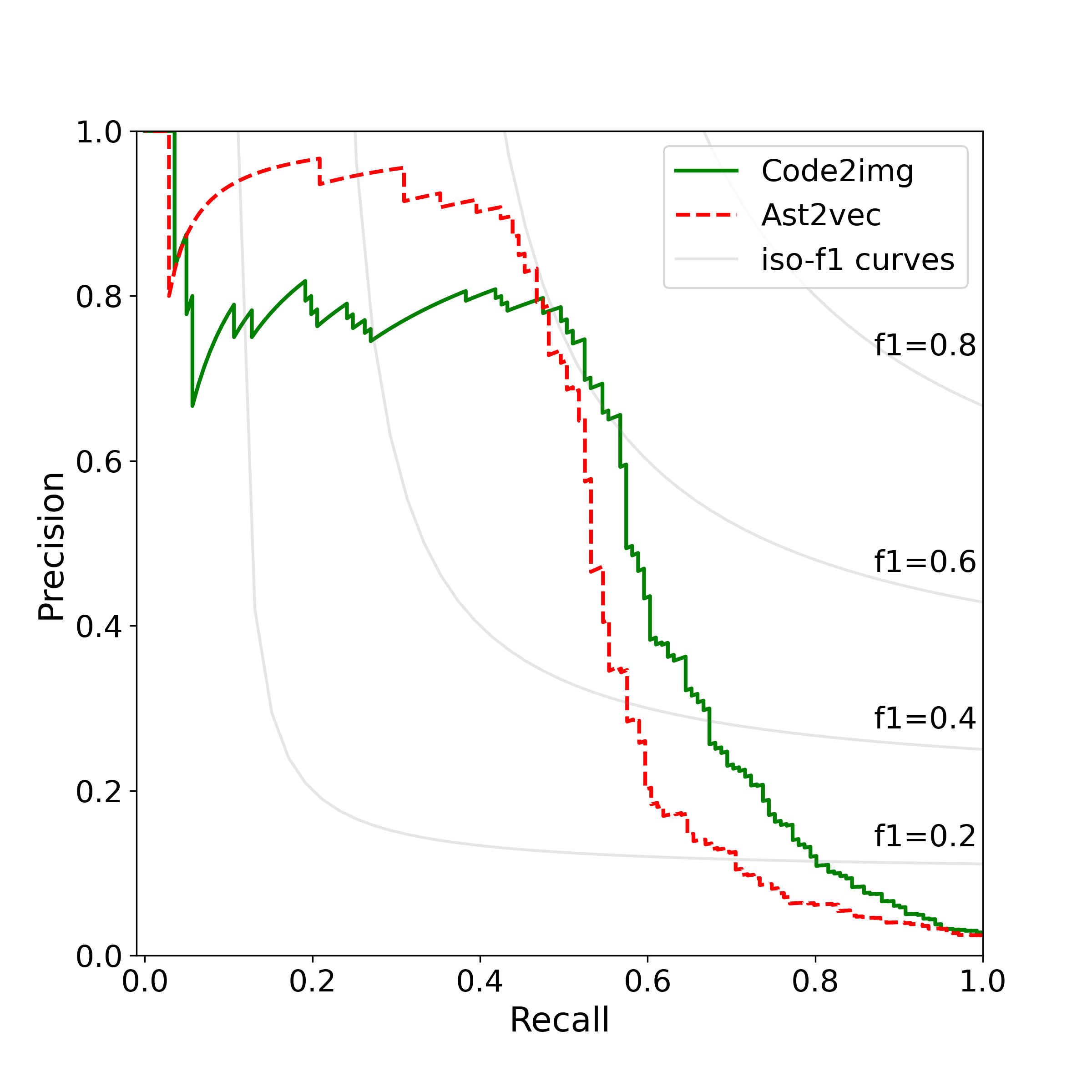}
  \caption{CWE-476}
  \label{fig:CWE476}
\end{subfigure}\hfil 
\begin{subfigure}{0.33\textwidth}
  \includegraphics[width=\linewidth]{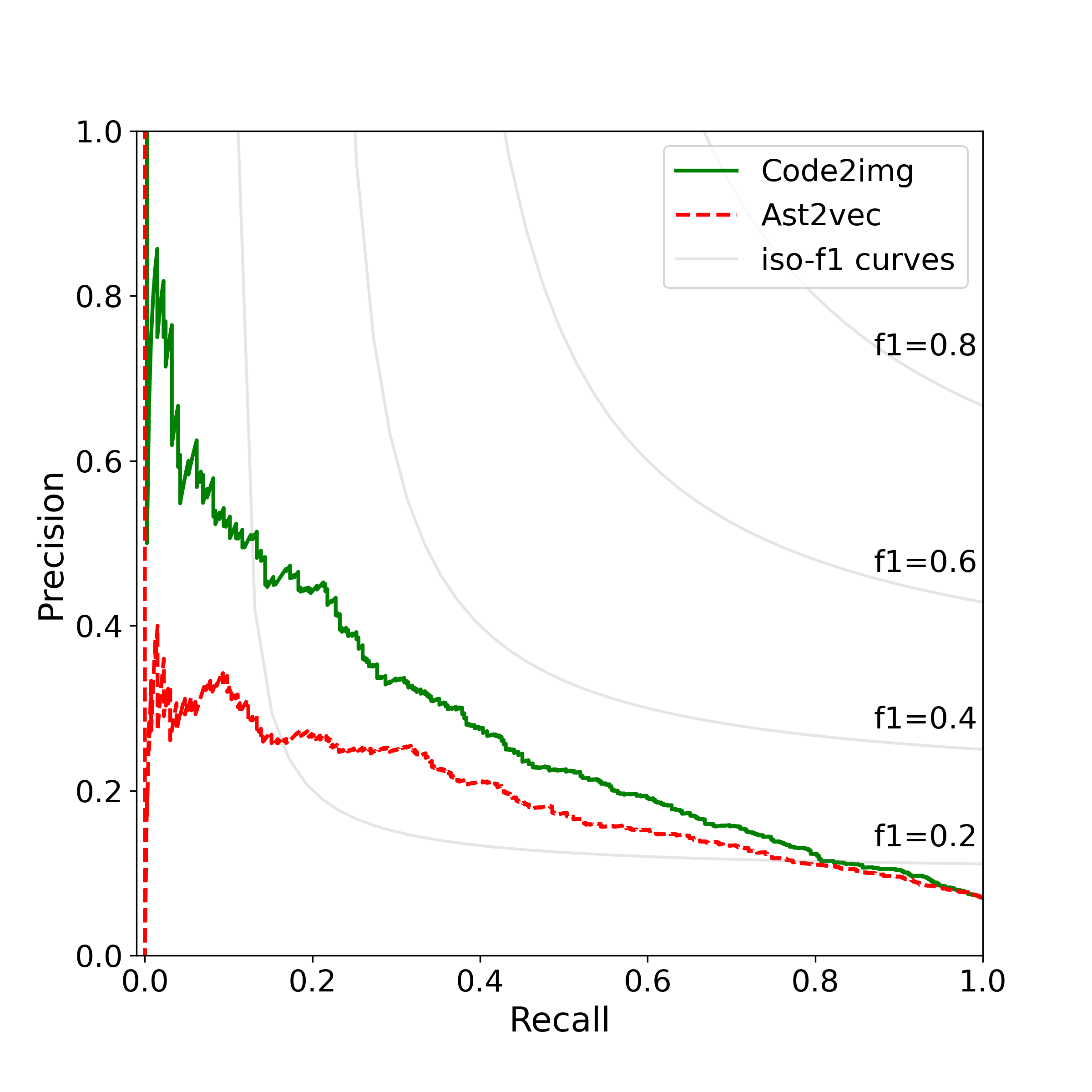}
  \caption{CWE-others}
  \label{fig:CWEothers}
\end{subfigure}
\caption{Performance comparison of the presented method with the Ast2vec \cite{Bilgin20}}
\label{fig:PRimages}
\end{figure*}

\begin{table*}[!h]
	\caption{Performance comparison of the presented method with the Ast2vec}
	\label{table-performancecompare}
	\centering
	\resizebox{\textwidth}{!}{
	\begin{tabular}{|l |c |c |c |c |c |c |c |c |c|c |c |c| c |c |c|}
		\hline
		\backslashbox{Method}{Vulnerability}
		& \multicolumn{3}{|c|}{CWE-119} & \multicolumn{3}{|c|}{CWE-120} & \multicolumn{3}{|c|}{CWE-469} & \multicolumn{3}{|c|}{CWE-476} & \multicolumn{3}{|c|}{CWE-others} \\
		\hline
		& F1 & MCC & AUC & F1 & MCC & AUC & F1 & MCC & AUC & F1 & MCC & AUC & F1 & MCC & AUC\\
		\hline
		Code2Image    & 0.668 & 0.648 & 0.554 & 0.528 & 0.463 & 0.477 & 0.140 & 0.166 & 0.054 & 0.617 & 0.613 & 0.524 & 0.335 & 0.282 & 0.277\\
		Ast2Vec \cite{Bilgin20} & 0.527 & 0.496 & 0.424 & 0.424 & 0.357 & 0.385 & 0.082 & 0.115 & 0.031 & 0.599 & 0.622 & 0.535 & 0.283 & 0.224 & 0.186\\
		\hline
	\end{tabular}}
\end{table*}

\textbf{CWE-119:} Figure \ref{fig:CWE119} shows precision-recall curves of both our Code2image method and the Ast2vec for vulnerability type of CWE-119, where it  obviously seems that the Code2image outperforms the Ast2vec. Indeed the Code2image has higher F1, MMC, and AUC scores than the Ast2vec as indicated in Table \ref{table-performancecompare}, where F1, MMC, and AUC scores  are 0.668, 0.648, and 0.554 respectively for the Code2image,  and 0.527, 0.496, and 0.424 respectively for the Ast2vec. The Code2image reaches its maximum F1 score of 0.668 at precision and recall values of 0.614 and 0.734 respectively.      

\textbf{CWE-120:} Figure \ref{fig:CWE120} depicts precision-recall curves of our Code2image method and the Ast2vec for the vulnerability group of CWE-120. As clearly seen in  Figure \ref{fig:CWE120}, the Code2image outperforms the  Ast2vec by reaching higher F1, MMC, and AUC scores. Table \ref{table-performancecompare} provides exact values of this metrics as 0.528, 0.463, 0.477 respectively for the Code2image, and as 0.424, 0.357, and 0.385 respectively for the  Ast2vec. The Code2image reaches its maximum F1 score of 0.528 when precision is 0.509 and recall is 0.549, while the Ast2vec reaches its maximum F1 score of 0.424 when precision is 0.373 and recall is 0.493.

\textbf{CWE-469:} Figure \ref{fig:CWE469} presents precision-recall curves for both compared methods. As can be inferred from Figure \ref{fig:CWE469}, predicting this vulnerability type is the most challenging one with respect to other categories because this category has the highest imbalance ratio on the test dataset. Indeed, it is seen in Table \ref{table-imbalanced} that there are only 32 positive samples 
 in the test dataset, making up approximately 0.56$\%$ of the whole test group. Still, our method shows better performance than the Ast2vec, with F1, MMC, and AUC values of 0.140, 0.166, and 0.054 respectively, whereas the same metrics for the Ast2vec are 0.082, 0.115, and 0.031 respectively, as given in Table \ref{table-performancecompare}.  The Code2image achieves its maximum F1 value of 0.140 when precision is 0.273 and recall is 0.094. The Ast2vec achieves its maximum F1 value of 0.082 when precision is 0.118 and recall is 0.063.

\textbf{CWE-476:} Figure \ref{fig:CWE476} depicts precision-recall curves for both  methods. It seems from  Figure \ref{fig:CWE476} that both methods performed similarly for this vulnerability type, and it is not clear whether one outperforms the other.  According to Table \ref{table-performancecompare}, there are slight differences in the F1, MMC and AUC scores of both methods such that they are  0.617, 0.613 and 0.524 respectively for the Code2image, and 
0.599, 0.622, and 0.535 respectively for the Ast2vec. The Code2image achieves its maximum F1 value of 0.617 when precision is 0.747 and recall is 0.525. The Ast2vec achieves its maximum F1 value of 0.599 when precision is 0.833 and recall is 0.468.  

\textbf{CWE-others:} Figure \ref{fig:CWEothers} exhibits precision-recall curves of both compared methods for last vulnerability category. It is obvious from Figure \ref{fig:CWEothers} that our Code2image method is superior to the Ast2vec. This is also seen  in Table \ref{table-performancecompare}, where the Code2image has F1, MMC, and AUC values of 0.335, 0.282, and 0.277 respectively, while the Ast2vec has  0.283, 0.224, and 0.186 for the same metrics respectively. The Code2image reaches its maximum F1 score of 0.335  when precision is 0.301 and recall is 0.379. The Ast2vec reaches its maximum F1 score of 0.283  when precision is 0.245 and recall is 0.335.

Based on the F1 scores given in Table \ref{table-performancecompare}, the Code2image showed its  best   performance for CWE-119, whereas the Ast2vec showed its best performance for CWE-476. Both methods performed worst for CWE-469 with respect to other vulnerability categories, which is probably due to fact that CWE-469 has the highest imbalance ratio that makes the classification task very challenging.

The Code2image shows better performance than the Ast2vec in general according to overall experimental results. One reason for this could be that the Ast2vec uses partial AST (not whole AST) for code representation due to some scalability issues, which causes information loss, while the Code2image generates code representation based on whole AST. Another reason is that we build a separate model for each vulnerability type in our method such that models are optimized to minimize validation loss, whereas a common model is built for all vulnerability categories in the Ast2vec.

\subsubsection{Inference Time}

We trained 5 different models, each of which was dedicated to a particular vulnerability type, up to the epoch number with the minimal validation loss as explained earlier. The models reached the minimal validation loss in different epochs (e.g. in epoch no 48 or 27),  which resulted in different training time for each model. But, the epoch-based training time was identical for all of them, which was observed as around 6 minutes per epoch on a machine with Nvidia GP106GL [Quadro P2000] GPU. On the other hand, we measured inference time by executing our trained models on the test dataset containing 5765 code samples, and found the average inference time per sample as 20.8 milliseconds.




\section{Conclusion}
\label{sec:conc}

We present a code representation method that converts AST of code to image form which can directly be entered into convolution neural networks. Thus it enables automatic feature extraction from source code by using deep learning algorithms. While generating such an image representation of code, syntactic and semantic information hidden in code is preserved thanks to AST. We implemented our method so that the size of the resulting image is optimized (i.e. compacted) as much as possible while retaining tree graph properties in visualization. For example, the lines connecting parent-child nodes should be drawn so that they do not intersect with each other except their common vertex if any. For validation, we realized a vulnerability prediction use case, and experimentally showed the superiority of our approach to state-of-art solutions.   

\bibliographystyle{plain}
\bibliography{references}
\appendix
\section{Additional examples}

In the following pages, we provide two more examples for source code, AST and corresponding visualised image representation generated by our code representation method. 
\begin{figure*}[h]
  \includegraphics[width=\textwidth]{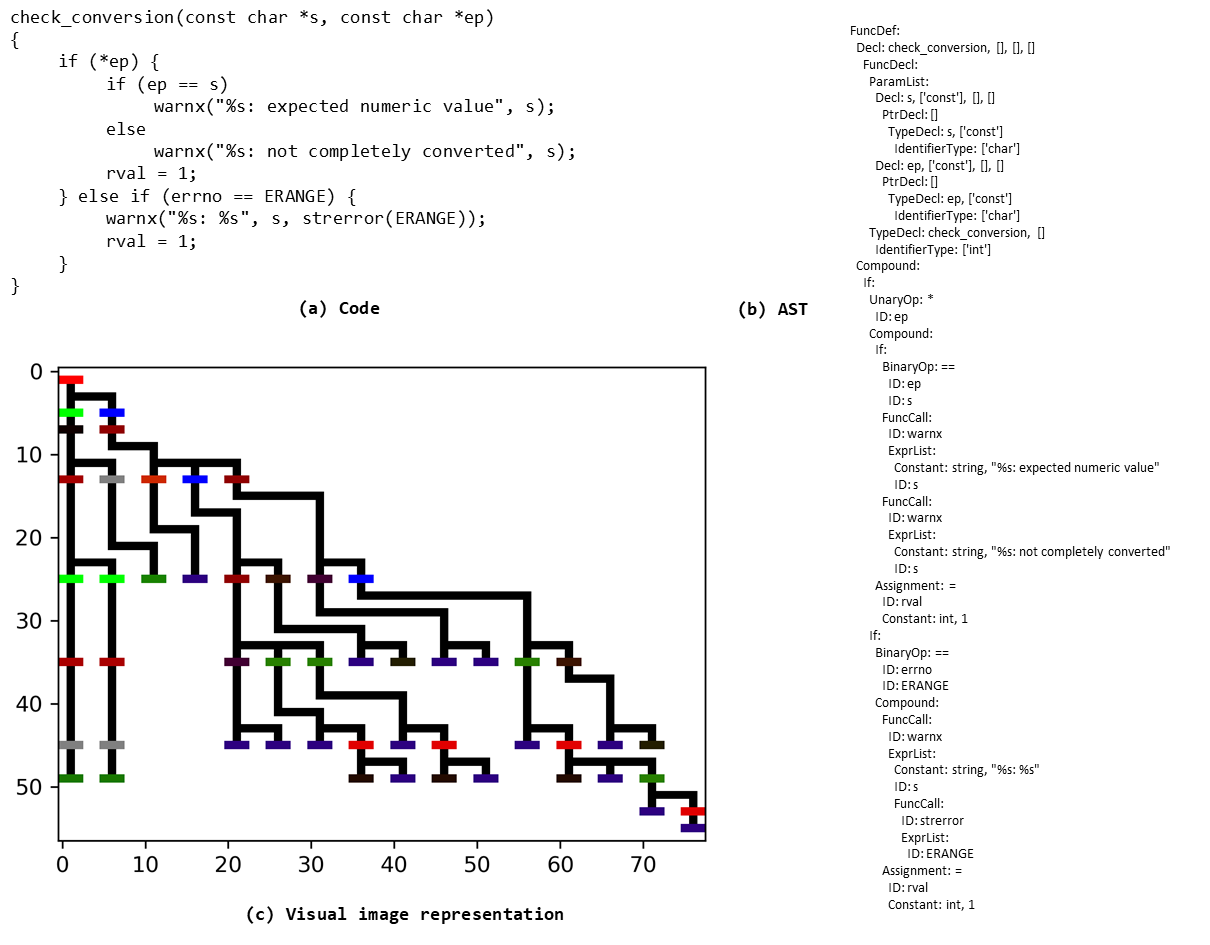}
  \caption{A real-world code sample from the Draper VDISC Dataset with its AST and corresponding visual image representation generated by our method. }
  \label{fig:appendixA}
\end{figure*}

\begin{figure*}[h]
    \centering 
  \includegraphics[width=\textwidth]{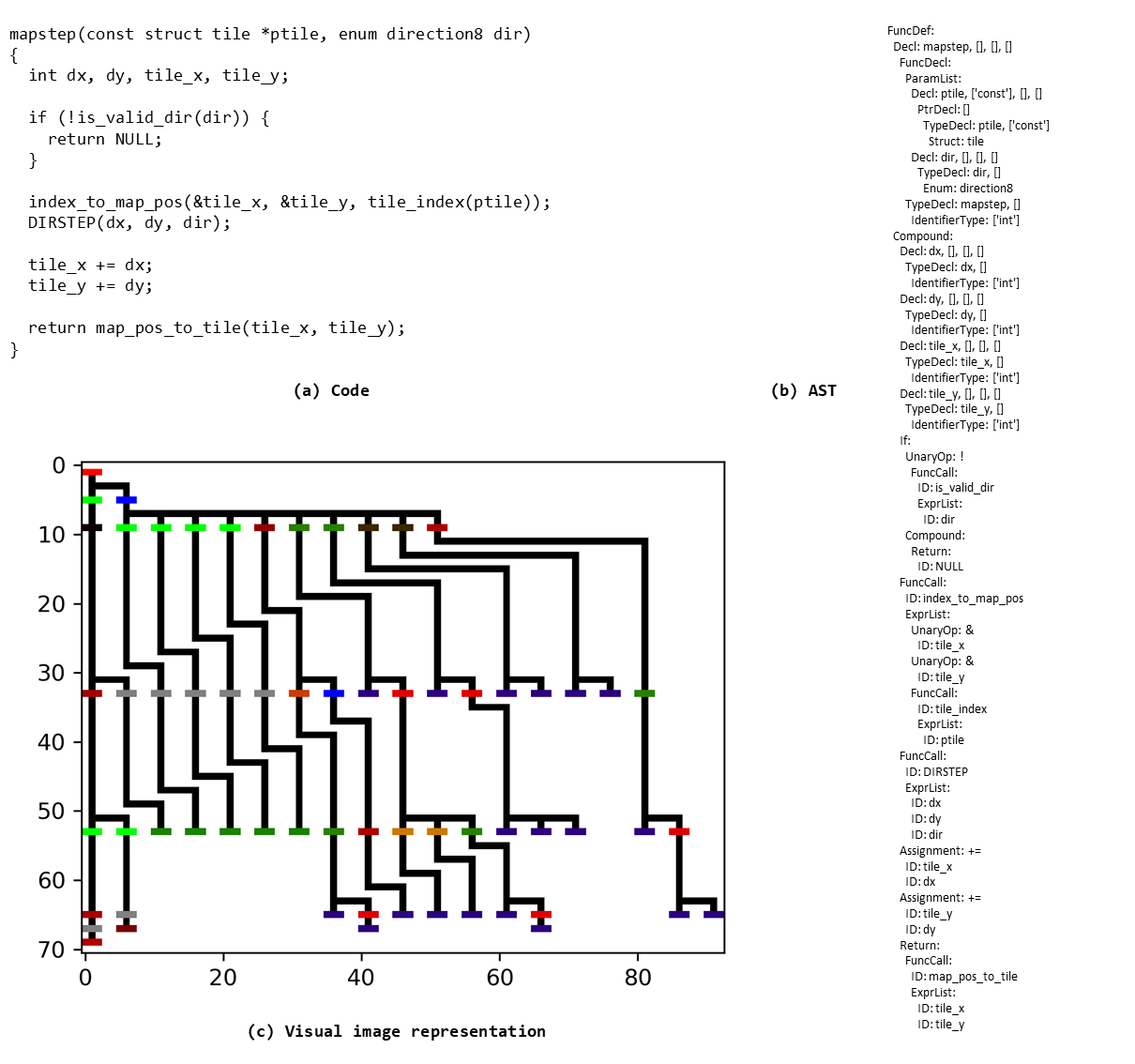}
  \caption{A real-world code sample from the Draper VDISC Dataset with its AST and corresponding visual image representation generated by our method.}
  \label{fig:appendixB}
\end{figure*}

\end{document}